\documentclass[12pt, preprint]{aastex}

\shorttitle{Investigation of the Star-Disk Interaction}
\shortauthors{Matt et al.}

\begin{document}

\title{Simulation-Based Investigation of a Model for the 
Interaction Between Stellar Magnetospheres and Circumstellar 
Accretion Disks}

\author{Sean Matt}
\affil{Astronomy Department, University of Washington, Seattle WA 98195}
\email{matt@astro.washington.edu}

\author{Anthony P. Goodson}
\affil{Earth \& Space Sciences, University of Washington, Seattle WA 98195}
\email{goodson@geophys.washington.edu}

\author{Robert M. Winglee}
\affil{Earth \& Space Sciences, University of Washington, Seattle WA 98195}
\email{winglee@geophys.washington.edu}

\and

\author{Karl-Heinz B\"ohm}
\affil{Astronomy Department, University of Washington, Seattle WA 98195}
\email{bohm@astro.washington.edu}

\begin{abstract}

We examine, parametrically, the interaction between the magnetosphere
of a rotating, young stellar object (YSO) and a circumstellar
accretion disk using 2.5-D (cylindrically symmetric) numerical
magnetoydrodynamic simulations.  The interaction drives a collimated
outflow, and we find that the jet formation mechanism is robust.  For
variations in initial disk density of a factor of 16, variations of
stellar dipole strength of a factor of 4, and for various initial
conditions with respect to the disk truncation radius and the
existence of a disk field, outflows with similar morphologies were
consistently produced.  Secondly, the system is self-regulating, where
the outflow properties depend relatively weakly on the parameters
above.  The large scale magnetic field structure rapidly evolves to a
configuration that removes angular momentum from the disk at a rate
that depends most strongly on the field and weakly on the rotation
rate of the foot-points of the field in the disk and the mass outflow
rate.  Third, the simulated jets are episodic, with the timescale of
jet outbursts identical to the timescale of magnetically induced
oscillations of the inner edge of the disk.  To better understand the
physics controlling these disk oscillations, we present a
semi-analytical model and confirm that the oscillation period is set
by the spin down rate of the disk inner edge.  Finally, our
simulations offer strong evidence that it is indeed the interaction of
the stellar magnetosphere with the disk, rather than some primordial
field in the disk itself, that is responsible for the formation of
jets from these systems.

\end{abstract}


\section{Introduction} \label{intro}

The formation of outflows from young stellar objects (YSO's) has been
a subject of interest for some time \citep[see summaries
by][]{koniglpudritz00,reipurthbally01}.  These outflows tend to
consist of a highly collimated ``jet'' and a poorly collimated
outflow, often referred to as the ``disk wind'' \citep{kwantademaru88,
kwantademaru95}.  Jets have outflow velocities of $\sim 50$ to 500 km
s$^{-1}$ and may, in some cases, exhibit spectral characteristics of a
hot plasma cooling down \citep{bacciotti97}.  The disk wind component
of the outflow typically has directionally averaged speeds of 0 -- 50
km s$^{-1}$ \citep{hirth3ea97, solfbohm93, solf97} and has spectral
characteristics of a cooler plasma.  Some systems contain a zero
velocity component, which may be due to an oblique shock interaction
between the disk wind and the disk.

As observational techniques and capabilities have improved, new
insights have been gained concerning the outflow properties very close
to the star.  For example, jet collimation can occur very close to the
source star-disk system.  Observations of DG Tauri using Solf's method
\citep{solf89} place the collimation region within 30 -- 40 AU, which
is the resolution of the observations of \citet{solfbohm93}.  Recent
measurements by \citet{bacciottiea00} of the spatial distribution of
the line emission for the H$\alpha$, [N II] 6583, [O I] 6363, and [S
II] 6731 lines, at different distances from DG Tauri, imply an even
closer collimation distance for this system of 14 AU.  Also, Hubble
Space Telescope observations of the HH 30 system \citep{burrowsea96}
imply that the jet must be fully collimated within 30 AU.  For a
general discussion of the collimation of YSO jets, see
\citet{eisloffelea00}.

Observations of HH 30 also imply that the jet formation mechanism is
episodic, producing outbursts (or ``knots'' along the jet axis) every
$\sim 2.5$ years \citep{burrowsea96}.  A probably related result is the
observation of DG Tau by \citet{solf97}, who showed that the
complicated [O I] 6300 line profile from the immediate environment of
DG Tau changes drastically within 1 year.  Larger scale observations
of other systems such as HH 111 imply a knot ejection timescale of
$\sim 15$ years \citep{reipurth3ea92}, but there is no consensus as to
whether or not the timescales of HH 30 and HH 111 are due to the same
physical principle.

One result of so many recent observational insights is that
significant progress has been made in understanding the mechanism
which generates YSO outflows.  While hydrodynamic means have been
investigated (e.g., a stellar wind which is collimated via shock
focusing; see, e.g., the early paper by \citealp{konigl82,
frankmellema96}), the models that have met the most success have
relied on magnetic fields to launch and collimate the outflow.  These
models fall into two categories: those that rely solely on a magnetic
field which threads the disk to centrifugally fling the wind (and
eventually collimate into a jet; \citealp[see,
e.g.,][]{pudritzouyed97, camenzind97, kuwabaraea00}), and those that
depend on the interaction between the stellar magnetic field and the
accretion disk \citep[e.g.,][]{lovelace3ea95, lovelace3ea99,
hayashi3ea96, millerstone97, goodson3ea97, goodson3ea99,
goodsonwinglee99}.  In the model of \citet[][henceforth
GWB]{goodson3ea97}, \citet[][henceforth GBW]{goodson3ea99}, and
\citet[][henceforth GW]{goodsonwinglee99}, a stellar dipole field
couples the star to the disk, differential rotation between the star
and the disk rapidly inflates the magnetic field via helicity
injection, and an outflow burst is produced with both jet and disk
wind properties.  Magnetic reconnection allows the process to repeat.
\citet{hayashi3ea96} and GWB identified this basic mechanism within a
few stellar radii, and GBW illustrated the resulting outflow out to
several AU with a magnetohydrodynamic (MHD) simulation in a single
case study.  GW laid out some of the physics driving key
characteristics of the outflow and star-disk interaction, allowing
some predictions to be made beyond their case study.

The work presented in GBW and GW implied that the interaction between
the stellar magnetosphere and the innermost regions of the accretion
disk could produce an outflow that is qualitatively similar to those
observed.  The outflow consists of a knotty jet (collimated within 1
AU) that is morphologically similar to the observations of HH 30
\citep{burrowsea96}.  However, the size scale is an order of magnitude
smaller than HH 30, and the timescale for knot formation is an order
of magnitude shorter.  In addition to the apparent mismatch between
observation and simulation results, issues that must be addressed are:

\begin{enumerate}

\item{The dipole field from the star initially threads the disk
everywhere.  Thus, it is not clear if the outflow is a result of the
interaction between the star and the disk (as claimed in GW), or if it
is due to the magnetic field which threads the disk everywhere at $t$
= 0.}

\item{The results of GBW represent a single case study. The 
sensitivity of the jet formation mechanism to key attributes of the 
star-disk system was not directly established.}

\end{enumerate}

To address these issues, we repeated the basic simulation of GBW for
various disk densities, stellar magnetic field strengths, and magnetic
field configurations.  Section \ref{simresults} contains the details
of the parametric cases and of the outflows that result in each case.
The parametric simulations also allow us to evaluate the disk
oscillation model of GW in \S \ref{evaluation}.  A semi-analytical
approach, including a formulation of the extraction of angular
momentum from the disk, leads to a modification and enables testing of
the modified model.  A summary and discussion follows, in \S
\ref{discussion}.

\section{Simulations and Outflow Results} \label{simresults}

We used the 2.5-D (axisymmetric) MHD simulation code employed by GWB,
GBW, and GW to examine, parametrically, the effect of varying the
accretion disk density, the stellar dipole field strength, and the
initial magnetic field configuration.  The simulation code employs a
two-step Lax-Wendroff (finite difference) scheme
\citep{richtmyermorton67} to simultaneously solve the following
time-dependent, MHD equations:
\begin{eqnarray}
{{\partial\rho}\over{\partial t}} &=& 
	-\mbox{\boldmath $\nabla$}\cdot(\rho \mbox{\boldmath $v$}) \\
\rho{{\partial(\mbox{\boldmath $v$})}\over\partial t} &=& 
	-\rho(\mbox{\boldmath $v$}\cdot\mbox{\boldmath $\nabla$})
	\mbox{\boldmath $v$}  -  \mbox{\boldmath $\nabla$}P
	-{{GM_*\rho}\over {(\mbox{\boldmath $r$}^2+\mbox{\boldmath $z$}^2)}}
	\hat{\mbox{\boldmath $R$}}
	+{{1}\over {\rm c}}(\mbox{\boldmath $J$}\times\mbox{\boldmath $B$}) 
	\label{mom_eq} \\
{{\partial e }\over\partial t} &=& 
	-\mbox{\boldmath $\nabla$}\cdot[\mbox{\boldmath $v$}(e + P)]
	-\left[{{GM_*\rho}\over 
	{(\mbox{\boldmath $r$}^2+\mbox{\boldmath $z$}^2)}}
	\hat{\mbox{\boldmath $R$}}\right]\cdot\mbox{\boldmath $v$}
	+\mbox{\boldmath $J$} \cdot \mbox{\boldmath $E$} \label{eng_eq}\\
{{\partial\mbox{\boldmath $B$}}\over\partial t} &=& 
	-{\rm c}(\mbox{\boldmath $\nabla$}\times\mbox{\boldmath $E$})
\end{eqnarray}
and uses
\begin{eqnarray}
\mbox{\boldmath $E$} &=& 
  -{{1}\over{\rm c}}(\mbox{\boldmath $v$}\times\mbox{\boldmath $B$})
	+\eta\mbox{\boldmath $J$} \label{efield_eq}\\
\mbox{\boldmath $J$} &=& 
  {{\rm c}\over {4\pi}}(\mbox{\boldmath $\nabla$}\times\mbox{\boldmath $B$}) \\
e 	&=& 
  {{1}\over {2}}\rho|\mbox{\boldmath $v$}|^2 + {{P}\over {\gamma - 1}}
\end{eqnarray}
where $\rho$ is the density, \mbox{\boldmath $v$} the velocity, $P$
the scalar gas pressure, $e$ the internal energy density,
\mbox{\boldmath $B$} the magnetic field, \mbox{\boldmath $J$} the
volume current, \mbox{\boldmath $E$} the electric field, c the speed
of light, and $\gamma$ the ratio of specific heats (we used $\gamma$ =
5/3 in our simulations).  In the momentum and energy equations
(\ref{mom_eq} and \ref{eng_eq}), we include a source term for gravity
in which $G$ is the gravitational constant, $M_*$ is the stellar mass,
$\mbox{\boldmath $r$}$ and $\mbox{\boldmath $z$}$ are the usual
cylindrical coordinates, and $\mbox{\boldmath $R$}$ is the spherical
distance from the center of the star ($|\mbox{\boldmath $R$}|^2 =
|\mbox{\boldmath $r$}|^2+|\mbox{\boldmath $z$}|^2$). The last term in
equation \ref{efield_eq} allows us to explicitly model the Ohmic
diffusion of the magnetic field by specifying the value of the
resistivity, $\eta$.  As discussed by GW, the numerical diffusivity
intrinsic to the simulations is measured to be $\approx 5 \times
10^{16}$ cm$^2$ s$^{-1}$.  In order to properly capture the physics of
magnetic diffusion, we use $\eta = 10^{17}$ cm$^2$ s$^{-1}$ in all of
our simulations.

We use a ``nested box'' scheme, and for the work presented here, all
simulations used 5 nested boxes, each with half the resolution (and so
twice the physical size) as the one interior to it.  Each box consists
of a 401 $\times$ 100 (in the cylindrical $r$ and $z$ direction,
respectively) grid.  As in GBW and GW, the grid spacing is 0.1
$R_\sun$ in the innermost box, but the outermost box dimensions are
3.2 AU by 0.8 AU.  The innermost (in $\mbox{\boldmath $R$}$) boundary
consists of a 15 gridpoint sphere (the ``surface of the star''), where
$\rho$, $P$, $\mbox{\boldmath $B$}$, and $\mbox{\boldmath $v$}$ are
held fixed at the initial values.  We enforce reflection symmetry
across the equatorial plane.  Along the rotation axis, we require that
$B_\phi = B_r = v_\phi = v_r = 0$, and all other quantities are equal
to the value at $r = dr$ (where $dr$ is the grid spacing).  We adopt
outflow conditions (the spatial derivative across the boundary is zero
for all quantities) at the outermost boundary of the outermost box.

The initial conditions for the baseline case of this study is
parametrically identical to the case illustrated by GBW.  They
describe the initial conditions for this case in considerable detail
(see their table 1), which includes a disk similar to a classic
$\alpha$ model accretion disk \citep{shakurasunyaev73} with $\dot
M_{\rm acc}$ = 10$^{-7}$ $M_\sun$ yr$^{-1}$ and $\alpha$ = 0.2
($r_{\rm au}$)$^{-1.5}$ (where $r_{\rm au}$ is the usual cylindrical
coordinate in units of AU).  The stellar magnetic field is dipolar,
with a field strength of 1800 G at the pole of the 1 $M_\odot$, 1.5
$R_\odot$ central star.  Initially, the stellar corona is in
hydrostatic equilibrium, and the density falls of as $\mbox{\boldmath
$R$}^{-4}$.  The star rotates with a period of 1.8 days (which is fairly
close to the shortest observed rotation period of CTTS; see, e.g.,
\citealp{bouvierea86}), and the disk is initially truncated at 8.5
$R_\odot$ and has a midplane density of $2.2 \times 10^{-10}$ g
cm$^{-3}$ at 10 $R_\odot$.  The dipole field threads the disk
everywhere at the start of the simulation.

To save computational time, this work focuses more on the outflow
generation mechanism and the integrated outflow properties instead of
the larger scale outflow morphology.  The simulation cases were all
carried out for a period of at least 160 days of real time (requiring
30-60 days of computing time per case).

Table \ref{summarytab} serves as a quick reference for all simulation
cases in this study.  The three cases which examine variations of
initial disk density are the ``low density,'' ``baseline,'' and ``high
density'' cases.  The three which examine variations of stellar field
strength are the ``weak field,'' ``baseline,'' and ``strong field''
cases.  The relative factors by which the disk density and the
magnetic field strength were varied were chosen based on energy
balance.  This is because the importance of the magnetic field is
measured by the ratio of the magnetic energy to the kinetic energy in
the disk.  Since the disk kinetic energy is proportional to the disk
density, and the magnetic energy scales as the square of the stellar
field strength, we varied the initial disk density by factors of 4 and
the initial magnetic field strength by factors of 2.

We also investigated simulations with a ``primordial,'' vertical
magnetic field in the disk.  So that the stellar magnetosphere was
truncated (at the region of disk and stellar field equality) near the
star, the simulations use stellar dipole strength equal to the weak
field case, plus a vertical field of 0.07 Gauss everywhere in the
simulation region.  These parameters give a stellar magnetosphere that
is truncated at 29 $R_\odot$ in the equatorial plane.  The stellar
magnetosphere has an open or closed topology, depending on the
polarity of the primordial field.  The vertical field in these
simulations cases is relatively weak (in that the stellar dipole is
dominant for an extended region; the ratio of magnetic to kinetic
energy near the disk inner edge is $\sim 10^{-5}$) but is sufficient
to assess the effect of the dipole field that initially threads the
disk at all radii in the other cases.  In order to initially exclude
the disk from the stellar magnetosphere, we truncated the disk at 29
$R_\odot$ for these models.  Therefore, our models with various
magnetic field topologies comprise three cases: One with a
downward-directed ($-z$) vertical field (``closed magnetosphere''),
one without a vertical field (``truncated disk''), and one with an
upward-directed field (``open magnetosphere'').  All three have a
stellar dipole field with the same strength as the weak field case,
and all have a disk truncation radius at 29 $R_\odot$ (which is 3.4
times further out than all the other simulations).

\subsection{Outflow Morphology}

Figures \ref{jetdensfig} -- \ref{jetgeomfig} illustrate the general
outflow morphology for all cases.  In these figures, the density
grayscale (logarithmic) and poloidal velocity vectors represent data
in the outermost box for a single ``snapshot'' in time.  Data from the
inner 100 $\times$ 100 grid points of the outer box are shown,
reflected about the rotation axis to better show the jet.  The
grayscale emphasizes material in the outflow, but does not capture the
relatively high density distribution near the star and disk inner
edge.  For a detailed look at the complex and time-dependent density
distribution near the star, see figures 3 and 4 of GBW.  Note that
different times are represented for each case.  This is because the
launching mechanism is essentially periodic, with the period being
different for each case.  The innermost regions of the disk oscillate
radially, and each oscillation triggers a large scale reconnection
event which alters the shape of the outflow in a periodic way.  For
the purpose of comparison, we show the various cases at a similar
phase (but not at the same time) of their outflow.

As evident in figure \ref{jetdensfig}, disk density variations of a
factor of 16 have no major effect on the outflow morphology.  The
basic outflow, in all cases, consists of a wide angle component (the
``disk wind'') and a jet.  Also note that even at this close-in scale
(on the order of an AU), the general characteristics of both the jets
and disk winds are similar -- knots in the jet are apparent in each
case (most clearly in the high-density case), as are the ``wispy''
structures between the disk and jet (associated with magnetic
reconnection; see GBW).

While the general morphologies are remarkably similar, some
differences between the outflows in figure \ref{jetdensfig} are
apparent.  The low density case has a higher outflow speed, a lower
outflow density, and a larger spacing between knots in the jet that
the other cases.  Conversely, the high density disk case has a higher
density, slower outflow, and the spacing between the knots in the jet
is reduced considerably.  This reduction in knot formation period is
related to a reduction in the period of radial disk oscillations
(discussed further in \S \ref{evaluation}).

Figure \ref{jetfieldfig} represents the simulation cases which
examined variations in stellar magnetic field strength.  Again, the
morphologies of all runs are similar, indicating that variations of a
factor of 4 in magnetic field strength produce no significant
variation in large scale morphology.  Each outflow produces a knotty
jet with a disk wind and a wispy structure in the region between the
disk and jet.  However, differences between the outflows in figure
\ref{jetfieldfig} are evident.  The outflow is slower in the weak
field case and faster in the strong field case.  Also, the structure
of the disk wind in the weak field case has a more complicated
appearance than the other two cases.

Figure \ref{jetgeomfig} illustrates the outflows for the cases with
different large scale magnetic field topologies.  Insets in each of
the three panels show the topology in the initial conditions.  Once
again, note that the general morphologies of the outflow are all
similar.  Recalling from table \ref{summarytab} that these three
topologies are variants of the weak field case, it is interesting to
note that all of these cases exhibit the same general characteristics
of the weak field case (fig.\ \ref{jetfieldfig}).  All produce
outflows with jets and disk winds (including wispy structures) with
velocities similar to the weak field case, and all three cases in
figure \ref{jetgeomfig} produce outflows with a complicated structure
(relative to the other cases) on the one AU scale.  This figure is
particularly noteworthy because, in the bottom two panels, the disk is
threaded everywhere by a vertical field, so it has much more magnetic
energy in the disk (at large radii) than the cases with only a stellar
dipole.  The fact that the outflows in these cases are not
significantly different from the top panel, supports the underlying
premise of the model: It is the interaction between the stellar
magnetosphere and the surrounding accretion disk that produces the
outflow, and the effect of the dipole field lines initially threading
the disk at large radii is insignificant.

The primary conclusion to be drawn from figures \ref{jetdensfig} -
\ref{jetgeomfig} is that the outflow formation mechanism is robust.
Also, to the extent these parameters were varied, the jet morphology
produced by this mechanism is independent of the density of the
accretion disk, the strength of the stellar magnetic field, and the
large scale magnetic field topology.  As seen in these figures,
however, the magnitude and timescales of outflow characteristics do
depend on the magnetic field strength and the density of the accretion
disk, and the remainder of this paper addresses these dependencies in
a quantitative way.

\subsection{Sensitivity of Outflow Properties \label{outflowprop}}

Figure \ref{maxvelfig} shows the variation in radial outflow
velocities for all cases.  For the data in this figure we first
calculated the radial velocity (${\mbox{\boldmath $v$}} \cdot \hat
{\mbox{\boldmath $R$}}$) on a sphere with a radius of 0.67 AU (90 grid
points in the outer box) centered on the star.  Due to the periodic
nature of the system, and because we were most interested in the
global outflow properties, we used a time average value.  Since it
takes a while for the outflow to clear out the entire simulation
region, our average only includes data from the interval of 50 -- 160
days.  We thus obtain the average velocity as a function of angle from
the pole.  The data in figure \ref{maxvelfig} represent the maximum of
the time-averaged velocity, which occurred at $\sim 40$ degrees for
all cases.  The leftmost panel shows the velocities as a function of
the disk variation, the middle as a function of the initial magnetic
field energy, and the right panel as a function of the vertical field
strength.  Note that the entire range of velocities (100 to 200 km
s$^{-1}$) is well within the range of observed outflow velocities (50
to 500 km s$^{-1}$) and corresponds to the velocities in most of the
observed HH objects \citep[see][]{raga3ea96}.

In panel a), the radial outflow velocity decreases with increasing
disk density, while in b), the outflow velocity increases with
increasing magnetic field energy.  Finally, panel c) shows that the
outflow velocity is roughly independent of the large-scale magnetic
field topology.  It is important to note that the velocities in panel
c) are similar to the weak field case.  Remember that the simulations
represented in c) have their disk initially truncated 3.3 times
further out than the weak field case, and two of them have an
additional vertical ``primordial'' field.  We see from panel b) that
the outflow velocity depends on the magnetic energy (from the stellar
dipole), so the fact that panel c) shows essentially no trend suggests
that it is only the magnetic energy in the stellar magnetosphere that
is important for accelerating the outflow.

The data in panels a) and b) can be fit with a power law ($v_{\rm w}
\propto \rho_{\rm d}^{n}$ and $v_{\rm w} \propto E_{\rm mag}^{n}$,
respectively, where $\rho_{\rm d}$ is the initial disk density and
$E_{\rm mag}$ is the initial magnetic energy associated with the
stellar dipole field).  A least squares minimization fit gives $n$ =
$-0.2$ and 0.3 for panel a) and b), respectively.  We note that there
is not necessarily a physical justification for fitting all of the
data with a power law, but we will use it to quantify the dependence
of various system properties on the parameters we varied.

Figure \ref{mdotfig} shows the variation in mass outflow rates, $\dot
M_{\rm w}$, for the three parametric studies.  To obtain these points
(and the ones for $L_{\rm w}$ and $\dot J_{\rm w}$, discussed below),
we first calculated the mass flux (also energy flux and angular
momentum flux) through a sphere at 0.67 AU.  The mass outflow rate
(also $L_{\rm w}$ and $\dot J_{\rm w}$) is given by an integration
over the surface of the sphere and a time average over the interval of
50 -- 160 days.  In an effort to exclude the disk (which can be
inflowing or outflowing, but with very high mass relative to the jet
and disk wind), the data shown represent an integration over the
surface up to an opening angle of 75 degrees.  To get an idea of the
relative uncertainties (due to the fact that we are representing a
time-dependent property by a time-averaged value) in the data, we
calculated the cumulative average of $\dot M_{\rm w}$ during the 50 --
160 day interval.  We take three times the standard deviation of this
cumulative average over the interval of 85 -- 160 days as our
uncertainty.  This quantity reflects the uncertainty due to the
periodicity of $\dot M_{\rm w}$ and due to the settling down of the
system to a fixed period (discussed further in \S \ref{evaluation}).

All of the data in figure \ref{mdotfig} have $\dot M_{\rm w} \sim
10^{-7}$ -- $10^{-8}$ $M_\sun$ yr$^{-1}$, consistent with observations
of YSO outflows \citep{hartmann97}.  Also, $\dot M_{\rm w}$ increases
with both disk density and magnetic field energy.  Once again, a power
law nicely quantifies this dependence with $n$ = 0.2 for both panel a)
and b) (within uncertainty).  The variation of mass outflow rate with
the different field topologies, panel c), shows no clear trend.

Figure \ref{luminosityfig} shows the variation in outflow
luminosity, $L_{\rm w}$, for the parametric cases studied.  The
energy flux is the sum of the kinetic, magnetic, and the thermal
energy fluxes.
\begin{equation}
\Phi_{\rm eng} = \left( {{1\over 2} \rho v^2 + {P\over {\gamma - 1}} 
+ {B^2 \over {8 \pi}}} \right) 
({\mbox{\boldmath $v$}} \cdot \hat {\mbox{\boldmath $R$}}),
\end{equation}
and the data in figure \ref{luminosityfig} represent the
time-averaged, integrated flux (as for $\dot M_{\rm w}$ above).  The
format of the figure is the same as figures \ref{maxvelfig} and
\ref{mdotfig}.  The disk density has no obvious influence on $L_{\rm
w}$, and the best fit power law is consistent with $n$ = 0.  The
outflow luminosity does, however, depend on the initial magnetic field
energy, suggesting a power law with $n$ = 0.4.  Finally, the addition
of a weak vertical field has no significant impact on outflow
luminosity.

An analysis of the angular momentum carried in the outflow, $\dot
J_{\rm w}$, is deferred to \S \ref{angmom}, where we use a
semi-analytical formulation to understand the controlling parameters
of the angular momentum extracted from the inner edge of the accretion
disk.

\subsection{Large Scale Magnetic Field Evolution \label{magfield}}

In the jet launching process described in GBW (also in
\citealp{hayashi3ea96}, and, to some degree, in
\citealp{lovelace3ea99}), the large scale magnetic field undergoes
rapid inflation.  This inflation results from the differential
rotation between the star and the disk which twists the magnetic
field, injecting helicity, and increasing the overall magnetic energy.
The resulting magnetic field has characteristics that differ
significantly from those of a dipole.  Figure \ref{expandfig}
illustrates the initial magnetic field expansion of the baseline case,
demonstrating the rapid change from a dipole to a more complicated and
open geometry.

The basic process of magnetospheric inflation can be summarized as
follows (see GWB and GBW for details): The star is initially connected
to the disk via the dipole magnetic field.  Differential rotation
between the star and disk induces an azimuthal component of the
magnetic field, increasing the magnetic pressure ($|{\mbox{\boldmath
$B$}}|^{2}/8\pi$) in the twisted region, which in turn inflates the
poloidal component of the magnetic field.

The expanded magnetic field has many properties that are different
from the simple dipole.  Figure \ref{bslicefig} illustrates the most
significant differences for the discussion at hand.  The figure shows
the total magnetic field strength, $|{\mbox{\boldmath $B$}}|$, on a 45
degree line in the meridional plane for the baseline case.  The lower
solid line shows the original dipole field.  The top solid line shows
the total field, $|{\mbox{\boldmath $B$}}|$(tot) = $(B_r^2 + B_\phi^2
+ B_z^2)^{1/2}$, and the middle solid line shows the poloidal field,
$|{\mbox{\boldmath $B$}}|$(pol) = $(B_r^2 + B_z^2)^{1/2}$, both at $t$
= 61.6 days.

Note that the dipole field decreases with distance as $R^{-3}$, the
evolved total field decreases as $R^{-n}$ with $n$ near 1.7, and the
evolved poloidal field decreases with $n$ roughly equal to 2.0.  Also
note that the evolved field has a ``dipole-like'' ($n \sim 3$) region
close to the star ($R \lesssim 5 R_\odot$), with a clear transition to
an ``expansion-dominated'' ($n \sim$ 1.5-2) region beyond.  The total
and poloidal field strengths are almost identical in the dipole-like
region, indicating that $B_\phi$ is negligible there.  A piece-wise
continuous fit was made of the expanded field (along a $45^\circ$
slice), of the form:
\begin{equation}
\label{twopowerfiteq}
|{\mbox{\boldmath $B$}}| = {\rm Max} \left[ B_{01} 
\left({R_* \over R}\right)^{n_{1}},B_{02} \left({R_* \over R}\right)^{n_{2}}
\right]
\end{equation}
In this fit, $n_1$ is the power law that applies in the interior,
dipole-like region, and $n_2$ is the power law that applies in the
exterior, expanded region.  $B_{01}$ and $B_{02}$ are the field
strengths that one would obtain at the surface of the star, if the
power law could be extrapolated all the way to the stellar surface
(this is roughly the true surface value for the dipole-like fit, but
not for the expanded field fit).  Both the total and poloidal field
are fit in this way.  The dashed lines in figure \ref{bslicefig} are
the fits to the total and poloidal field at $t = 61.6$ days.

Table \ref{bpowertab} lists the values of the magnetic field fit
parameters for all of the simulations.  Since the values are
time-dependent (periodic), and it takes some time for the flow to be
established in the entire simulation grid, table \ref{bpowertab} data
have been averaged during the interval of 50 -- 160 days.  To get an
idea of the uncertainty of these values (due to the episodic nature of
the system), we first obtain the cumulative average during this
interval.  The uncertainty listed in the table is the standard
deviation of this cumulative average during the interval of 85 -- 160
days.  For each case, column 2 and 3 contain the $B_{01}$ (in Gauss)
and $n_1$ values.  Since the total and poloidal field is nearly
identical in this interior fit, we list only the value for the
poloidal field.  The last 4 columns contain similar values for the
exterior portion of the fit for both the total and poloidal field
components.

When $n_2$(pol) is near 2.0, the poloidal field is essentially open
($n = 2$ for a purely radial field).  This is roughly true for all
cases, though the ones with weaker dipole fields (row 4 and 6 -- 8)
have $n_2$(pol) significantly less than than 2.  The primary
conclusion from this data is that the total magnetic field strength
will behave as $R^{-n}$ where $n$ is between 1.5 and 2.0.  The cases
with additional vertical field in the disk (last two rows of table
\ref{bpowertab}) have the largest magnetic to kinetic energy ratios
(far from the disk inner edge) of all cases.  However, since the
magnetic field in the outflow originates in the stellar magnetosphere,
the field in the evolved outflow of these cases is identical to the
weak field and truncated disk case.

\subsection{Summary of Parametric Results}

The most striking result of this parametric study is that every case
examined produces an outflow similar to that of GBW.  For a variation
of disk density of a factor of 16, a variation in stellar magnetic
field of a factor of 4, and variations in initial disk truncation
points and large scale magnetic field topologies, all of the
simulations produce outflows with jets and disk winds.  The only
differences are in the details of the star-disk interaction and in the
magnitude of the outflow properties.

Table \ref{sensitivetab} summarizes the sensitivities of the outflow
properties and morphology to increasing the disk density and
increasing the stellar magnetic energy.  The listed numerical values
are the spectral index ($n$) from power laws that represent various
data (e.g., the dotted lines in figures \ref{maxvelfig} -
\ref{luminosityfig}).  The least squares power law fits were weighted
by the uncertainty (as described in \S \ref{outflowprop}), and the
errors in table \ref{sensitivetab} are the formal 1-sigma errors of
the fit.  Note that some of outflow properties listed in table
\ref{sensitivetab} are discussed in the following section.

The main result of table \ref{sensitivetab} is that variation of key
system parameters do indeed cause variations in outflow properties,
but the relationship is weak (i.e., $|n|$ is small).  The
star-disk-field system is self-regulating, due to the interaction
between the disk and the outflow.  The system adjusts to mitigate the
impact of any single parameter change.  Also, the large scale field
topology does not significantly affect any of the outflow properties.
This result implies that it is indeed the interaction between the
stellar magnetosphere and the accretion disk that drives the outflow,
as opposed to the existence (via artificial initial conditions as in
GBW) of a dipole field that threads the disk everywhere at $t$ = 0.

\section{Evaluation of the Disk Oscillation Model of GW 
\label{evaluation}}

As noted in GBW and GW, the inner edge of the disk undergoes radial
oscillations.  These oscillations are actually a flux exchange process
-- on each oscillation, the stellar magnetosphere diffuses into the
innermost regions of the disk and differential rotation inflates the
magnetic field to an open configuration.  The inflated field topology
spins the disk down via magnetocentrifugal launching of disk plasma
\citep[see, e.g.,][]{blandfordpayne82}.  As it spins down and moves
inward toward the star, the disk inner edge is stripped away from the
rest of the disk.  At the point of ``closest approach'', a
reconnection event alters the magnetic field topology, and the spun
down portion of the disk flows along the newly reconnected field
lines, accreting onto the star.

Figure \ref{oscillationfig} illustrates the disk oscillations for the
simulation cases where the accretion disk density is varied.  The
solid line is the radial location of the inner edge of the disk as a
function of time.  The dashed line shows the location of the Keplerian
co-rotation point (where the stellar solid body rotation rate is equal
to the Keplerian rotation rate).  The figure clearly demonstrates the
dependence of disk oscillations on the disk density.  As disk density
is decreased, the amplitude and period of the oscillations increase.
Note that it takes time for the system to settle down to a constant
period (the physical explanation for this is discussed in \S
\ref{oscillation}).  In most cases, the system becomes quasi-periodic
after about 50 days, but the low density case seems only to be
settling down near the end of the 160 day simulation.  Much of our
uncertainty in all quantities is due to this effect, and we have
calculated our uncertainties to reflect this.

As the average disk location approaches the Keplerian co-rotation
point, the difference between the disk and stellar magnetosphere
rotation rates is reduced, so the magnetosphere will expand more
slowly.  This is true in the high density case (bottom panel), but
here the disk oscillations are actually faster than in the other
cases.  Also, even in the high density case, the oscillation period is
much longer than the inflation timescale (which is roughly the time
for the disk inner edge and star to rotate to a differential angle of
$\pi$; see \citealp{uzdenskyea02}; GW).  This supports the basic tenet of GW
that the oscillation period is driven by the spin-down rate of the
inner edge of the disk (via magnetic torques) rather than the
inflation rate of the magnetic field.

Ideally, the parametric simulation results presented here would allow
for the disk oscillation model of GW to be tested.  Unfortunately, the
data in \S \ref{magfield} require us to relax their assumption that
the expanded magnetic field has reached a power law configuration
characterized by $|{\mbox{\boldmath $B$}}| = B_*(R_*/R)^{2}$.  While
this may be true for the poloidal field of many cases (see table
\ref{bpowertab}), it is more generally true that the expanded field
reaches a configuration with $|{\mbox{\boldmath $B$}}| =
B_0(R_*/R)^{n}$, with $n$ between 1.5 and 2.0 (and $B_0$ not
necessarily equal to the stellar surface value).  We will also discard
the assumption of GW that the accretion disk has reached a secular
steady state, where the accretion rate onto the star (regulated by the
interaction between the stellar magnetosphere and the inner edge of
the disk) equals the accretion rate given by the $\alpha$-disk
description of the disk.  While this secular steady state does exist
near the end of simulations, the accretion rate in the outer portion
of the simulated disk is unlikely to be equal to the $\alpha$-disk
model.  This is because the physics of the Shakura-Sunyaev
$\alpha$-disk model are not fully captured in the simulations.
However, we will show in the following section that the extraction of
angular momentum from the inner edge of the disk can be simply
understood and determined from simulation data.  In light of this, we
present a modified disk oscillation model (in \S \ref{oscillation}),
in which the angular momentum extraction can be a prescribed value --
thus removing the assumption of secular stead state in the disk.

We should point out that it is our ultimate goal to be able to
determine the properties of the outflow ($\dot M_{\rm w}$, $v_{\rm
w}$, $L_{\rm w}$, oscillation period, etc.) generated by the star-disk
interaction a priori from the physical parameters of the system
($R_*$, $B_*$, disk accretion rate, etc.).  Unfortunately, the
complexity of the model, which includes a relatively large number of
parameters, makes this difficult.  The following work in this section
is thus meant to be an early step in that direction.  As such, this
semi-analytical work relies on some quantities derived from the
outcome of the simulations (e.g., $B_0$).  The work presented in \S
\ref{angmom} and \S \ref{oscillation}, therefore, is useful for
testing our understanding of the most relevant physics controlling the
oscillations, and thus the accretion/ejection mechanism.  It will also
serve as a check for self-consistency in the simulations.

\subsection{Extraction of Angular Momentum from the Disk
\label{angmom}}


Consider an inner region of the disk that is threaded by a magnetic
field.  From the analytical work of \citet{weberdavis67}, we note that
the angular momentum transported in the wind of a magnetic rotator is
$\dot J_{\rm w} = \dot M_{\rm w} \Omega_{\rm fp} R_{\rm A}^2$ (where
$\Omega_{\rm fp}$ is the angular velocity of the footpoint of the flux
tube, and $R_{\rm A}$ is the Alfv\`en radius).  Assuming nearly rigid
rotation of plasma out to $R_{\rm A}$ (for a discussion of this
assumption, see, e.g., \citealp{michel69}), plasma will be
magnetocentrifugally launched along field lines via the
Blandford-Payne mechanism \citep{blandfordpayne82}, provided the field
lines make an angle of more than 30 degrees from vertical.  This angle
is typically 45 to 55 degrees in our simulations.  For a simple
approach, we'll assume that the velocity of the wind at the Alfv\`en
radius is $v_{R_{\rm A}} = \Omega_{\rm fp} R_{\rm A}$.  This is
strictly true for $R_{\rm A} \gg R_{\rm d}$ (where $R_{\rm d}$ is the
position of the field footpoint in the disk).  The Alfv\`en radius,
then, can be estimated by examining the energy balance at that
location:
\begin{equation}
\label{dfgs}
{1 \over 2} \rho_{R_{\rm A}} v_{R_{\rm A}}^2 = {B_{R_{\rm A}}^2 \over
{8\pi}},
\end{equation}
When expressed in terms of outflow properties, and assuming the
magnetic field behaves as $|{\mbox{\boldmath $B$}}| = B_0(R_*/R)^{n}$,
this becomes
\begin{equation}
\label{engybalouteq}
{1 \over 2} {{\dot M_{\rm w} \Omega_{\rm fp}} \over 
{4\pi f R_{\rm A}}} = {B_0^2 \over {8\pi}} 
\left( {R_* \over R_{\rm A}} \right)^{2n}
\end{equation}
Here, we have assumed that $\dot M_{\rm w} = 4 \pi R_{\rm A}^2
\rho_{R_{\rm A}} v_{R_{\rm A}}$.  Note that this assumption is not
valid for an anisotropic wind, so we introduced a filling factor, $f$,
by which the mass outflow rate is reduced.  Reducing equation
\ref{engybalouteq} and solving for $R_{\rm A}$ gives
\begin{equation}
\label{raeq}
R_{\rm A} = \left( {f B_0^2 R_*^{2n}} \over 
{\dot M_{\rm w} \Omega_{\rm fp}} \right)^{1 \over 2n-1}
\end{equation}
and so
\begin{equation}
\label{jdoteq}
\dot J_{\rm w} = (\dot M_{\rm w} \Omega_{\rm fp})^{1-{2\over 2n-1}}
(f B_0^2 R_*^{2n})^{2\over 2n-1}
\end{equation}

The dependence of $\dot J_{\rm w}$ on various outflow properties is
interesting.  The strength of the dependence relies only on $n$, the
magnetic field power law fall off.  Note that if $n$ = 1.5, equation
\ref{jdoteq} becomes
\begin{equation}
\label{jdot2eq}
\dot J_{\rm w} = f B_0^2 R_*^3,
\end{equation}
and $\dot J_{\rm w}$ is independent of both $\dot M_{\rm w}$ and
$\Omega_{\rm fp}$.  In general, for any $n$ between 1.5 and 2.0, the
exponent on the factor $(\dot M_{\rm w} \Omega_{\rm fp})$ in equation
\ref{jdoteq} is substantially smaller than on the factor $(f B_0^2
R_*^{2n})$.  Therefore, the extraction of angular momentum should be
most sensitive to the magnetic field strength.

The simulations verify this conclusion.  Figure \ref{jcompfig} shows
the angular momentum outflow rate for all cases.  The data (squares)
are calculated as described in \S \ref{outflowprop}, and shown in the
same format as figure \ref{maxvelfig}.  It is clear that the angular
momentum carried in the simulation outflow is more dependent on
magnetic energy than on disk density, which influences $\dot M_{\rm
w}$ and $\Omega_{\rm fp}$ (see figures \ref{mdotfig} and
\ref{oscillationfig}).  In fact, panel a) of figure \ref{jcompfig}
shows no obvious dependence of $\dot J_{\rm w}$ on disk density at all
(also see the fifth row of table \ref{sensitivetab}).  This also
appears to be true for the cases with various field topologies shown
in panel c), where there is no clear trend, especially considering the
larger uncertainties.

The simulations also offer a quantitative test of equation
\ref{jdoteq}.  We used the time-averaged simulation values for $\dot
M_{\rm w}$, $\Omega_{\rm fp}$, $B_0$, and $n$ (the magnetic field
power law fall off) for each case and calculated $\dot J_{\rm w}$
using equation \ref{jdoteq}.  Since the inner edge of the disk spends
most of the time beyond 5 $R_\odot$ (see figure \ref{oscillationfig}),
the value of $B_0$ and $n$ to use comes from the outer portion of the
piece-wise fit to the magnetic field strength in the simulations.
Also, since angular momentum is added to the wind via the poloidal
field anchored in the disk, the calculated values of $\dot J_{\rm w}$
use $B_{02}$(pol) and $n_2$(pol).  We chose a single value of the
filling factor ($f$ = 0.43), so that the calculated values match the
baseline result.  These calculated values (triangles in figure
\ref{jcompfig}) can be directly compared to the simulated values
(squares), and the agreement between them is remarkable.  The largest
disagreement is (for the closed magnetosphere case) at the 50\% level.

In the context of equation \ref{jdoteq} and the assumptions at the
beginning of this section, it seems clear that variations in the
location of the inner edge of the disk (which is the launching point
for much of the outflow) and in $\dot M_{\rm w}$ (which is related to
the conditions at the footpoint of the flux tube) have little relative
impact on the angular momentum ejected from the system.  The primary
reason for this is the restructuring of the initial magnetic field.
As evident in table \ref{bpowertab} (also see figs \ref{expandfig} and
\ref{bslicefig}), the magnetic field rapidly evolves into a state that
follows a radial power law with $n$ between 1.5 and 2.0, increasing
the importance of $B_0$ relative to $\dot M_{\rm w}$ and $\Omega_{\rm
fp}$.

In addition to this, the magnetic field in the outflow seems to be
restructured in such a way that it ``softens'' the effect (on $\dot
J_{\rm w}$) of the parameters varied.  For example, the $B_{02}$(tot)
values for the weak field, baseline, and strong field cases are 550,
750, and 1180 Gauss, respectively.  The entire range in these values
is only a factor of 2, even though the entire range of stellar surface
dipole field strengths (which is the only simulation parameter changed
among these cases) is a factor of 4.  Thus, a given variation in the
stellar surface magnetic field strength cannot be used in conjunction
with equation \ref{jdoteq} alone to determine the resulting variation
in $\dot J_{\rm w}$.  The self-regulating nature of the system indeed
makes prediction difficult.

\subsection{Modified Disk Oscillation Model \label{oscillation}}

In their oscillation model, GW assumed that the magnetic field
associated with the stellar magnetosphere would diffuse into the disk
radially at a given radial velocity ($v_{\rm diff}$).  The differing
azimuthal velocities of the inner edge of the disk and the stellar
magnetosphere (which rotates at the angular rotation rate of the star)
leads to a rapid inflation of the magnetosphere (note that $v_\phi \gg
v_r$ and $v_{\rm diff}$).  The inflated magnetic configuration removes
angular momentum from the inner edge of the disk (via
magnetocentrifugal launching of disk plasma), spinning it down.  The
inner edge of the disk will eventually achieve the condition where its
inward radial velocity exceeds the velocity at which the magnetic
field is diffusing out into the disk ($|v_r| \gtrsim |v_{\rm diff}|$).
At this point, the inner edge of the disk separates from the rest of
the disk, and spirals inward to the star, where a fraction of it is
ultimately accreted by the star, and a fraction is expelled in the
outflow.  Thus, the diffusive velocity ($v_{\rm diff}$) determines how
much of the disk separates for each oscillation, and so sets the mass
accretion rate.

There are two main assumptions in the disk oscillation model of GW
that must be modified.  They calculated the radial velocity of the
disk inner edge (as it spins down; their eq.\ 9) as a function of $r$,
$v_{\rm diff}$, $R_{\rm d}$ (the radial position of the disk inner
edge), $\Sigma(R_{\rm d})$ (the surface density there), and $B_*$.  We
first modify their assumption that $|{\mbox{\boldmath $B$}}| =
B_*(R_*/R)^2$ to the more general case, $|{\mbox{\boldmath $B$}}| =
B_0(R_*/R)^{n}$, and get
\begin{equation}
\label{diskmotioneq}
v_r = - B_0^2 R_*^{2n} r^{2-2n} 
{\{
G M_* [2 \pi R_{\rm d} \Sigma(R_{\rm d})]^2 r - 
B_0^2 R_*^{2n} \pi R_{\rm d} \Sigma(R_{\rm d}) r^{4-2n}\}^{1\over 2}
\over
G M_* [2 \pi R_{\rm d} \Sigma(R_{\rm d})]^2 -
B_0^2 R_*^{2n} \pi R_{\rm d} \Sigma(R_{\rm d}) (4-2n) r^{3-2n}}
\end{equation}
Note that when $n$ = 2, equation \ref{diskmotioneq} reduces to GW's
equation 9.  

The second assumption of GW to discard is that the accretion rate from
the disk inner edge (regulated by the interaction of star's
magnetosphere and the disk) is the same as the accretion rate given
analytically by the $\alpha$-disk description.  Instead, we assume that
the angular momentum carried in the outflow (see \S \ref{angmom}
and fig.\ \ref{jcompfig}) is equal to the amount extracted from the
disk inner edge.

Figures \ref{amplitudefig} and \ref{periodfig} contain the simulated
disk oscillation amplitudes and periods (squares and dotted lines; in
the same format as fig.\ \ref{maxvelfig}).  Since it takes some time
for the oscillations to settle down (see fig.\ \ref{oscillationfig}),
the simulated amplitude of the disk oscillations plotted in figure
\ref{amplitudefig} are the maximum separation of the star and disk
inner edge 160 days into the simulation.  The simulated periods
(fig.\ \ref{periodfig}) are also taken at this time.  As a measure of
the uncertainty (represented by error bars in the figures), we use the
difference between the amplitude at 160 days and the value at 85 days.
For the period, we use the same method, but the error bars have been
doubled to show up better on the plot.

The ``settling down'' of the disk oscillations is a result of the
initial conditions.  Equation \ref{diskmotioneq} indicates that, as
$R_{\rm d}$ increases, $v_r$ decreases (due to a decrease in the
magnetic torque and an increase of the amount of angular momentum at
that radius), and a ring from the disk inner edge has further to fall,
so the period increases.  Also, GW showed (their fig.\ 7) that the
accretion rate decreases for increasing $R_{\rm d}$.  If the initial
state of the system (in the simulations) is such that the accretion
rate regulated by the star-disk interaction ($\dot M_{\rm inner}$) is
greater than the accretion rate further out in the disk (at $r >
R_{\rm d}$; $\dot M_{\rm outer}$), the inner region of the disk will
be cleared out (thus increasing $R_{\rm d}$) until the two accretion
rates are equal.

Notice that the cases in panels a) of figures \ref{amplitudefig} and
\ref{periodfig} show a relatively strong variation in amplitude and
period, even though the angular momentum extraction rate is roughly
equal among them (see fig.\ \ref{jcompfig}).  This is explained by the
fact that, for a denser disk, $\dot M_{\rm outer}$ is larger, so the
final state of the system will have a smaller amplitude and,
therefore, a shorter period.

The previous section showed that angular momentum extraction depends
most strongly on the magnetic field strength.  This seems to imply
that, for stronger magnetic fields, the oscillation periods should be
shorter (since a ring of disk material would spiral in faster).  This
would indeed be the case if $R_{\rm d}$ were a fixed quantity.
However, when the magnetosphere-disk interaction can extract more
angular momentum from a given radius, $\dot M_{\rm inner}$ increases,
so the final state of the system will actually have a larger $R_{\rm
d}$ and a longer period.  This is evident in panels b) of figures
\ref{jcompfig} -- \ref{periodfig}.

To calculate the period and amplitude of disk oscillations, following
GW, we used an iterative process in which we first pick a radial
location for the disk inner edge ($R_{\rm d}$) and integrate equation
\ref{diskmotioneq} to obtain a spin down time.  Equation
\ref{diskmotioneq} requires the case-specific values of $B_{02}$(tot)
and $n_2$(tot) listed in table \ref{bpowertab}.  This method gives a
simultaneous prediction of the oscillation amplitude and period.  As
opposed to GW, we adopt the $R_{\rm d}$ value that satisfies the
requirement that the angular momentum extracted from the ring of
material (as it falls toward the star) divided by the oscillation
period (taken as twice the spin down time) equals the angular momentum
extraction rate in the simulations (displayed in fig.\
\ref{jcompfig}).  As in GW, we need an estimate of $v_{\rm diff}$, in
order to calculate the mass of the ring of material that is stripped
off of the disk inner edge.  We use $v_{\rm diff} \approx \eta/\beta
h$ ($h$ is the initial disk scale height and depends on $R_{\rm d}$),
which differs from GW's approximation by the factor $\beta$.  This
factor is necessary because the average disk scale height during the
simulations is larger than the initial value.  This has two causes: a)
the magnetic interaction increases the scale height above the initial
analytical value, and b) the numerical diffusion of disk material is
not negligible ($h$ is typically only a few grid points at $R_{\rm
d}$).  For each case, then, we used our explit value of $\eta =
10^{17}$ cm$^2$ s$^{-1}$, and chose $\beta$ to best fit the disk
oscillation period.  In all cases, $\beta$ ranged from 3 to 9.

The predictions of the modified model presented here (triangles in
fig.'s \ref{amplitudefig} and \ref{periodfig}) match the simulated
results (squares) well for the cases with various stellar magnetic
field strengths and disk densities, but are consistently too small for
the cases with various field topologies.  Remember that the the cases
in panel c) of figures \ref{amplitudefig} and \ref{periodfig} have a
disk which is initially truncated further out than all other cases.
Since their initial truncation radius is further out than the final
disk oscillation amplitude of the weak field case (to which they are
otherwise similar), these cases have an initial state in which $\dot
M_{\rm inner} < \dot M_{\rm outer}$.  The mismatch between the data an
predictions for these models may be due to a significant alteration of
the disk density profile from its initial state.

The agreement of the predictions of this modified model with the data
implies that the fundamental assumptions of the model are valid (at
least in the simulations).  Those assumptions are: 1) The disk
oscillations are associated with a flux interchange process where
magnetic flux diffuses into the inner-most edge of the disk.  2) The
disk oscillations are driven by spin-down torques associated with the
magnetic field that threads the disk.  3) The spin-down rate is driven
by the surface density of the accretion disk and the magnetic field
topology.  4) What appear to be large scale disk oscillation are
actually the spin-down of a ring of material from the inner most edge
of the disk.

\section{Summary and Discussion} \label{discussion}

The mechanism presented by GBW and GW for the launching of jets is
promising, but it is difficult to make a priori predictions, given the
large number of parameters in the star-disk system.  One has to
consider the effects of the mass of the star and disk, the magnetic
field strength, and the stellar rotation rate, to name a few.  In
addition, the disk responds to the conditions in its environment, so
one must also consider the self-consistent interaction between the
disk, outflow, and changing magnetic field topology.  The complexity
of this model prompted us to carry out a parameter survey using
magnetohydrodynamic simulations.  Using the case presented by GBW as a
baseline, we examined cases with various initial disk densities,
various stellar dipole field strengths, and cases with additional
``primordial'' field associated with the disk.  The results of this
survey are:

\begin{enumerate}

\item{The fundamental mechanism for jet formation is robust.}

\item{The star-disk-outflow system is self-regulating.}

\item{The mechanism is independent of magnetic flux that may exist in
the disk.}

\item{We have improved the disk oscillation model of GW.}

\end{enumerate}

\begin{figure}
\epsscale{0.5}
\plotone{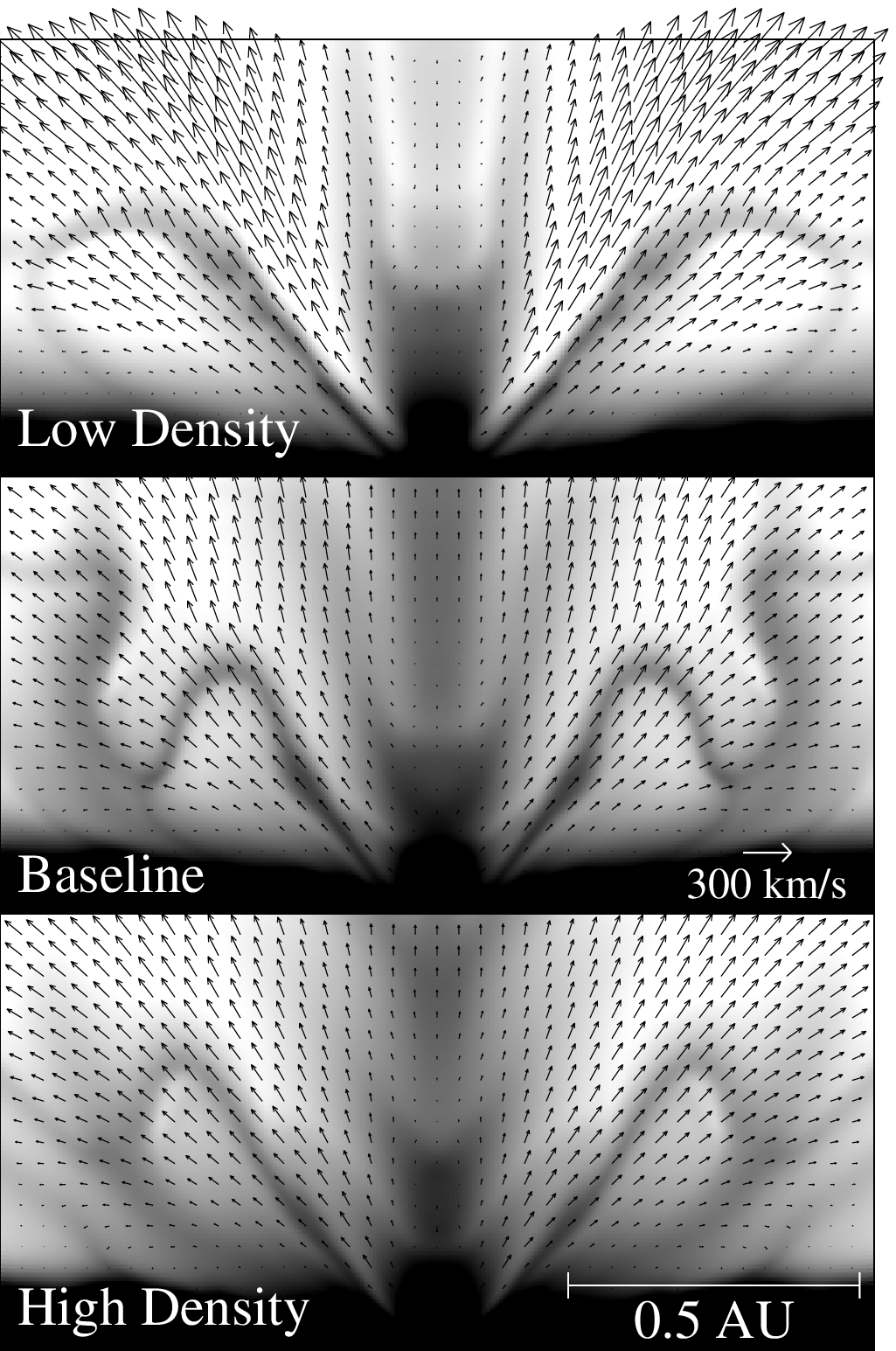}

\caption{A snapshot of density and poloidal velocity vectors
illustrates the jet and disk wind for all cases of disk density
variation. Grayscale ($\log n \leq 10^{-20.7}$, for $n$ in cm$^{-3}$,
is white and $\geq 10^{-18.5}$ is black) and velocity scale is
consistent for all frames.  The panels, from top to bottom, are the
low density case (at $t = 147$ days), the baseline case (at $t =
136^{\rm d}$), and the high density case (at $t = 137^{\rm d}$).  In
each panel, the equatorial plane is along the bottom, and the star is
at bottom center.  \label{jetdensfig}}

\end{figure}

\begin{figure}
\plotone{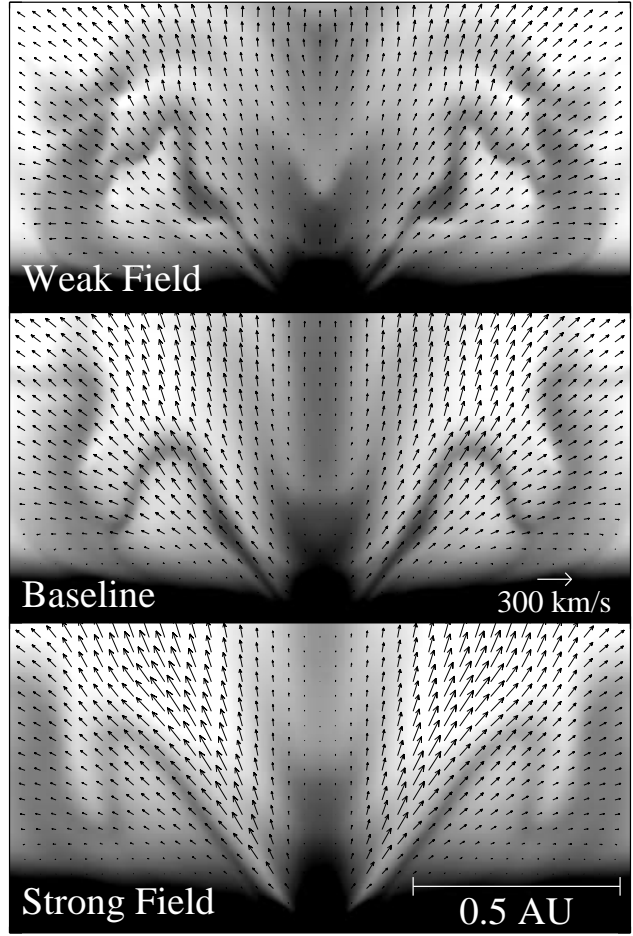}

\caption{Same as figure \ref{jetdensfig} but for weak field (at $t =
137^{\rm d}$), baseline (at $t = 136^{\rm d}$), and strong field (at
$t = 129^{\rm d}$) cases. \label{jetfieldfig}}

\end{figure}

\begin{figure}
\plotone{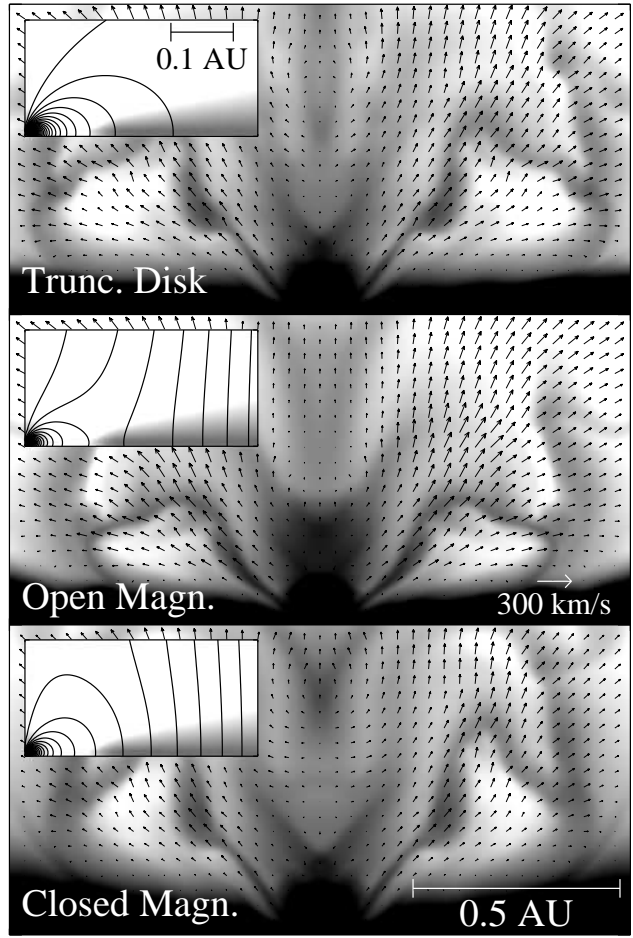}
\epsscale{1.0}

\caption{Same as figure \ref{jetdensfig} but for truncated disk (at $t
= 143^{\rm d}$), open magnetosphere (at $t = 138^{\rm d}$), and closed
magnetosphere cases (at $t = 145^{\rm d}$).  The insets in each panel
show the magnetic field topologies at the start of the simulation (at
a smaller scale; star is at lower left). \label{jetgeomfig}}

\end{figure}

The robust nature of the mechanism is best demonstrated by examination
of figures \ref{jetdensfig} -- \ref{jetgeomfig}.  For variations of
initial disk density, variations in stellar magnetic field strength,
and variations in initial disk truncation points and large scale
magnetic field topologies, all of the simulations produced outflows
with similar morphologies.  All outflows consisted of a highly
collimated, knotty jet and a less collimated ``disk wind'' (including
a ``wispy'' structure associated with magnetic reconnection).  Also,
the outflows were always episodic with the period set by the
oscillations of the inner edge of the disk.  The magnitude and
timescales of outflow characteristics, however, do depend on the
magnetic field strength and the density of the accretion disk, and we
were able to address these dependencies in a quantitative way.

\begin{figure*}
\plotone{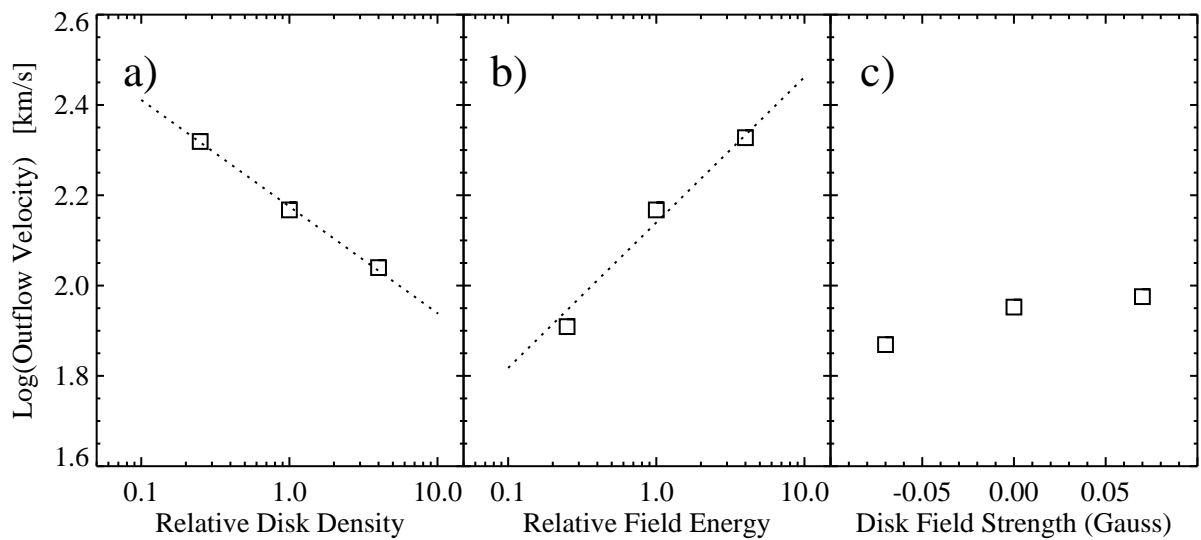}

\caption{The maximum radial outflow velocities (see text for details),
for all cases, reveal a dependence on system parameters.  The panels
show variation with respect to a) disk density, b) stellar magnetic
field energy, and c) the strength of a vertical magnetic field (in
addition to the stellar dipole field).  The abscissa values in panels
a) and b) are relative to the baseline case.  The dotted lines are
power law fits to the data. \label{maxvelfig}}

\end{figure*}

\begin{figure*}
\plotone{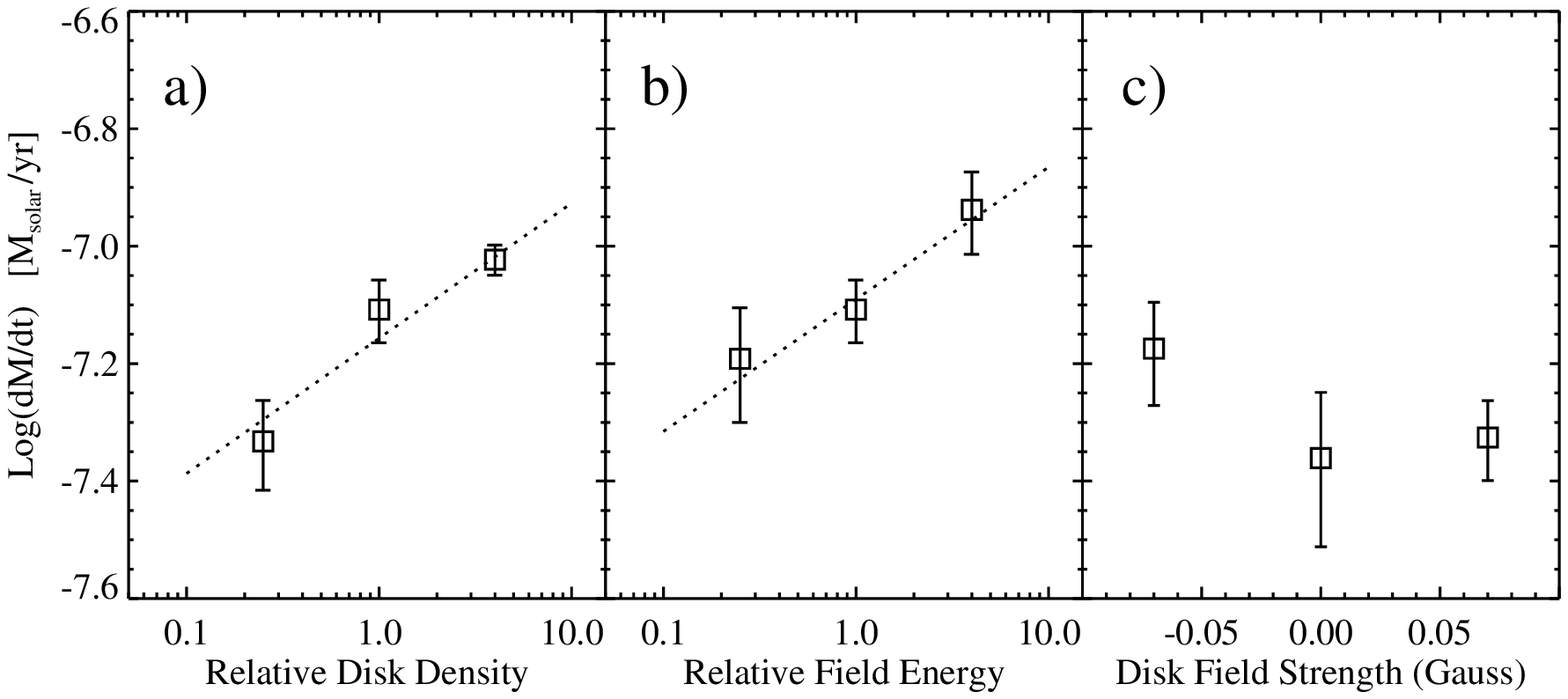}

\caption{Similar to figure \ref{maxvelfig} but for mass outflow
rates, $\dot M_{\rm w}$. \label{mdotfig}}

\end{figure*}

\begin{figure*}
\plotone{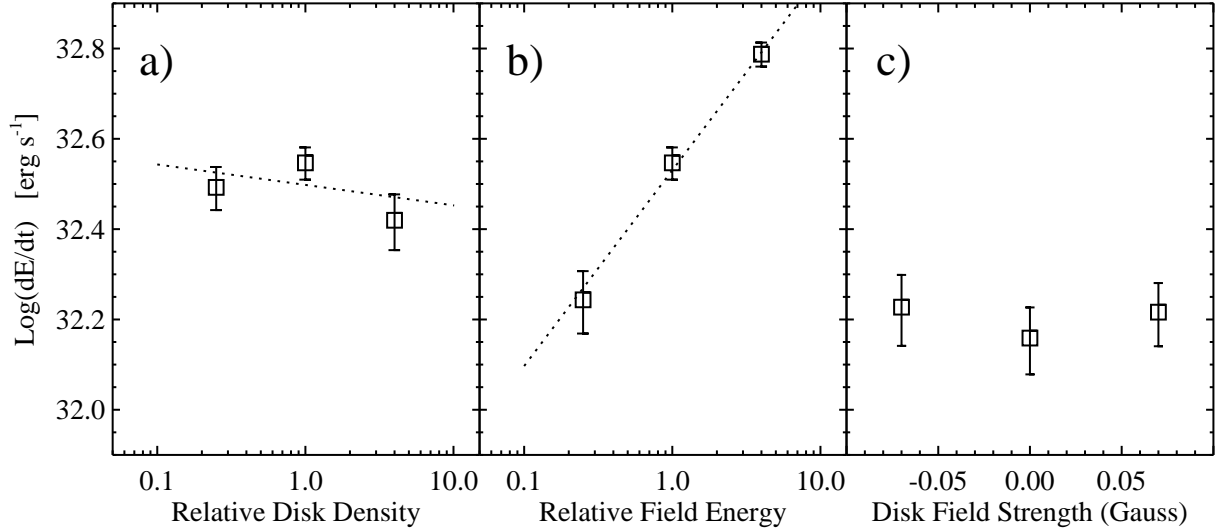}

\caption{Similar to figure \ref{maxvelfig} but for outflow luminosity,
$L_{\rm w}$ (see text). \label{luminosityfig}}

\end{figure*}

\begin{figure}
\epsscale{0.5}
\plotone{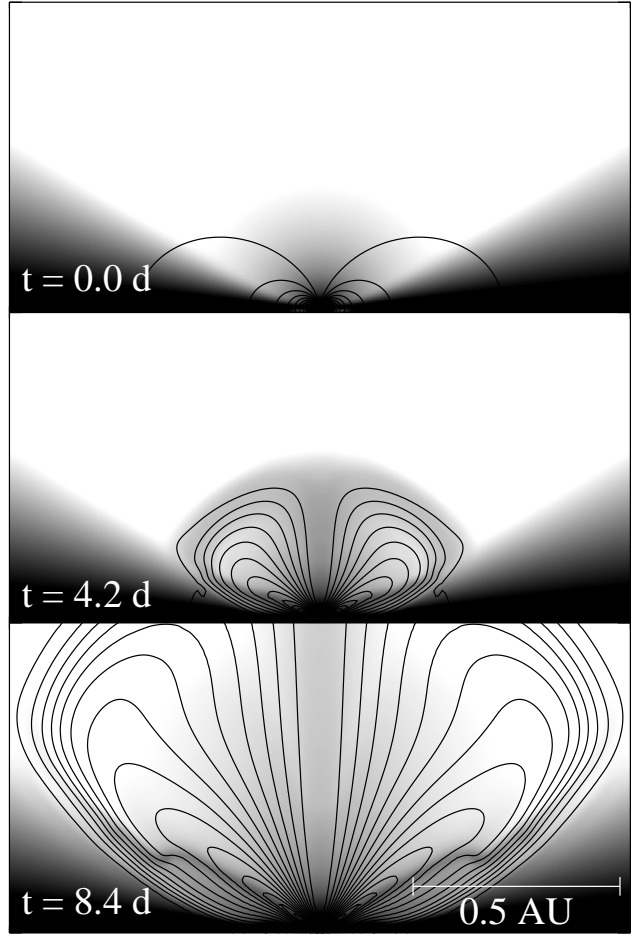}
\epsscale{1.0}

\caption{Snapshots from different times, early in the evolution of the
baseline case, illustrate the rapid expansion of the poloidal magnetic
field (solid lines).  The grayscale is $\log n$ ($\leq 10^{-22}$ is
white and $\geq 10^{-17.5}$ is black, for $n$ in
cm$^{-3}$). \label{expandfig}}

\end{figure}

\begin{figure}
\plotone{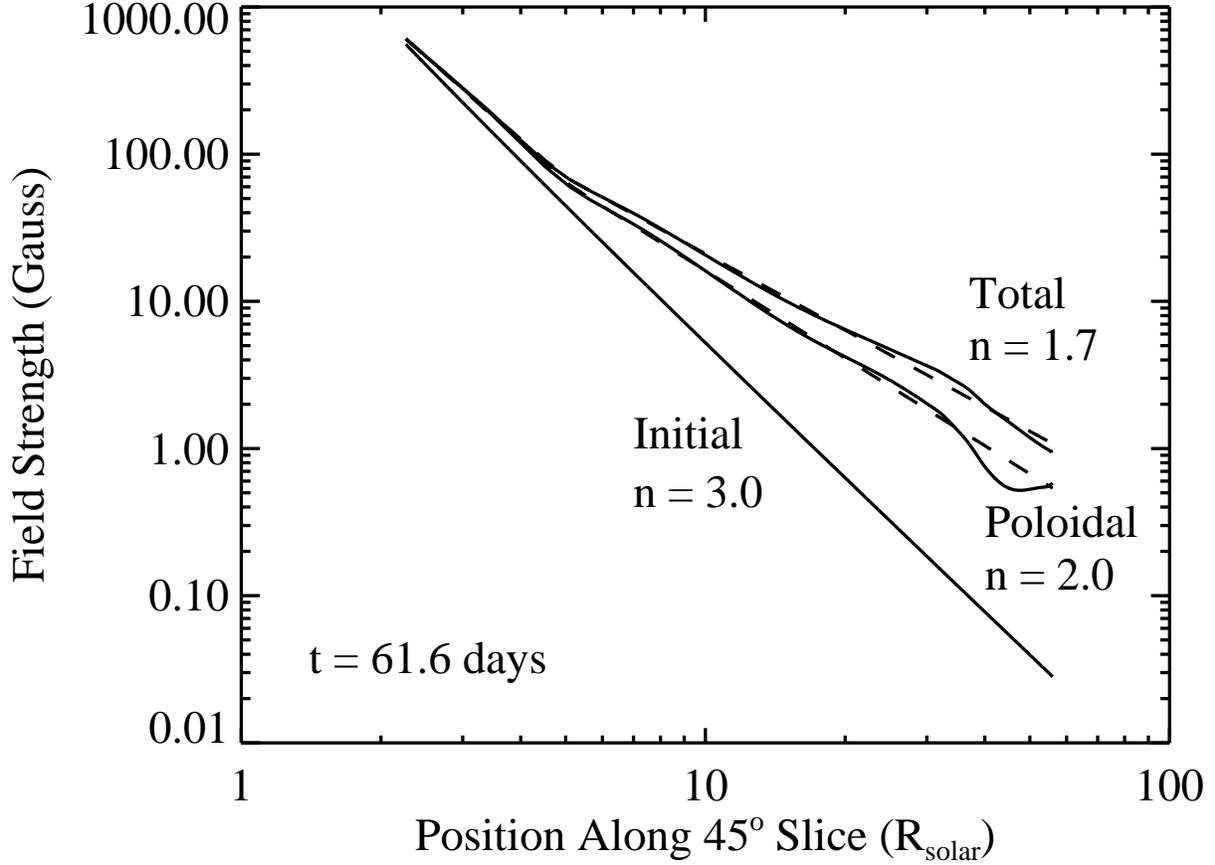}

\caption{The magnetic field strength as a function of distance from
the central star (along a 45 degree line in the meridional plane)
quantifies the expansion of the magnetic field for the baseline case.
The lower solid line is the initial dipole field, the top solid line
is the strength of the evolved total magnetic field at $t$ = 61.6
days, and the middle solid line is the poloidal field at the same
time.  Dashed lines are broken power law fits to the data, and the
power, $n$, in the exterior region of the fit is
shown. \label{bslicefig}}

\end{figure}

\begin{figure}
\epsscale{0.5}
\plotone{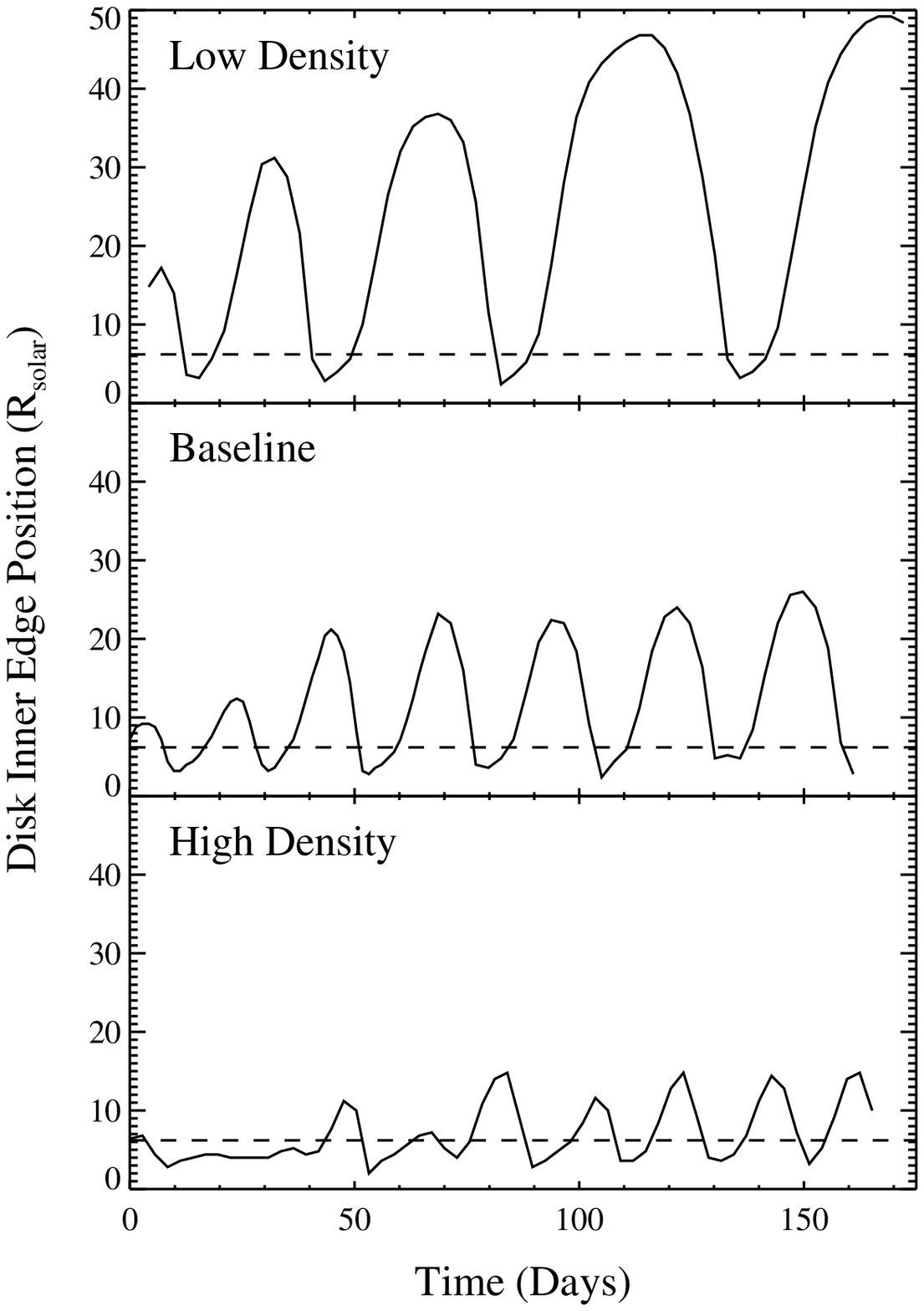}
\epsscale{1.0}

\caption{The radial location of the inner edge of the accretion disk
as a function of time (solid line) is sensitive to variations in the
disk mass.  The three panels correspond to three cases with different
initial disk densities, and the dashed line in each is the location of
the Keplerian co-rotation point.  \label{oscillationfig}}

\end{figure}

\begin{figure*}
\plotone{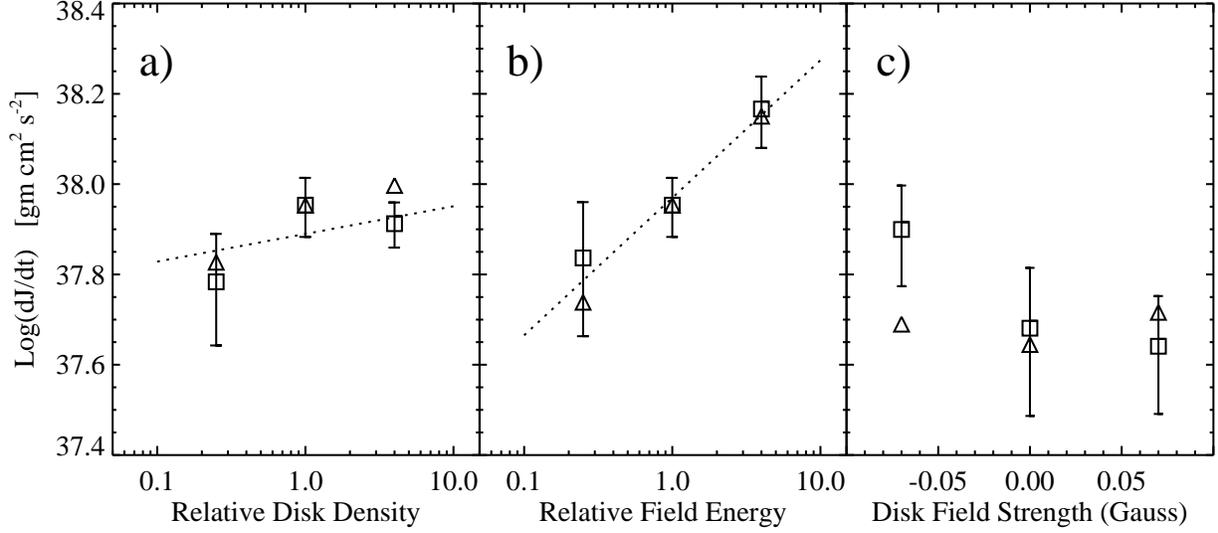}

\caption{The squares and dotted lines are as in figure \ref{maxvelfig}
but for angular momentum carried in the outflow, $\dot J_{\rm w}$.
The triangles represent a semi-analytical calculation for each case
(see text).
\label{jcompfig}}

\end{figure*}

\begin{figure*}
\plotone{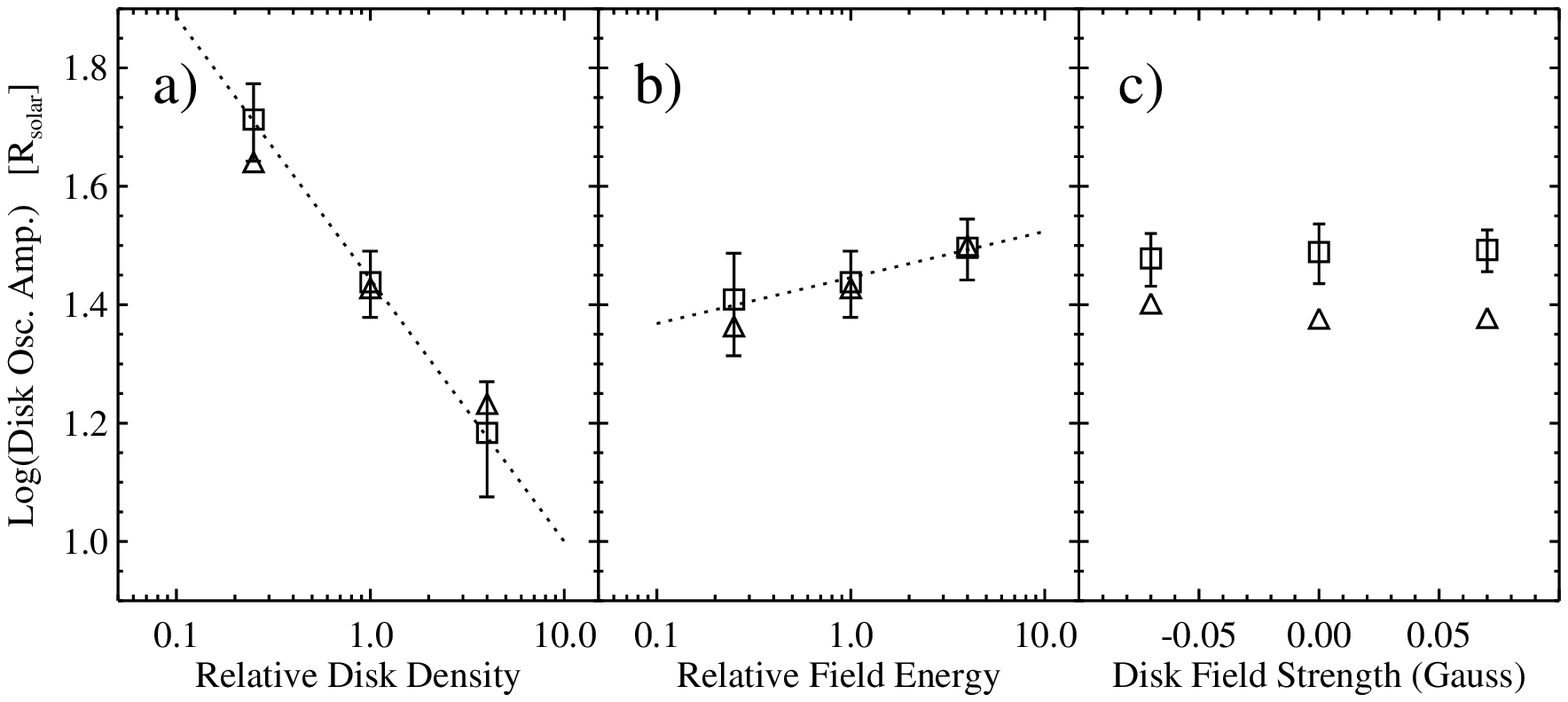}

\caption{The squares and dotted lines are as in figure \ref{maxvelfig}
but for disk oscillation amplitudes at $t$ = 160 days.  The triangles
represent a semi-analytical calculation (see text).
\label{amplitudefig}}

\end{figure*}

\begin{figure*}
\plotone{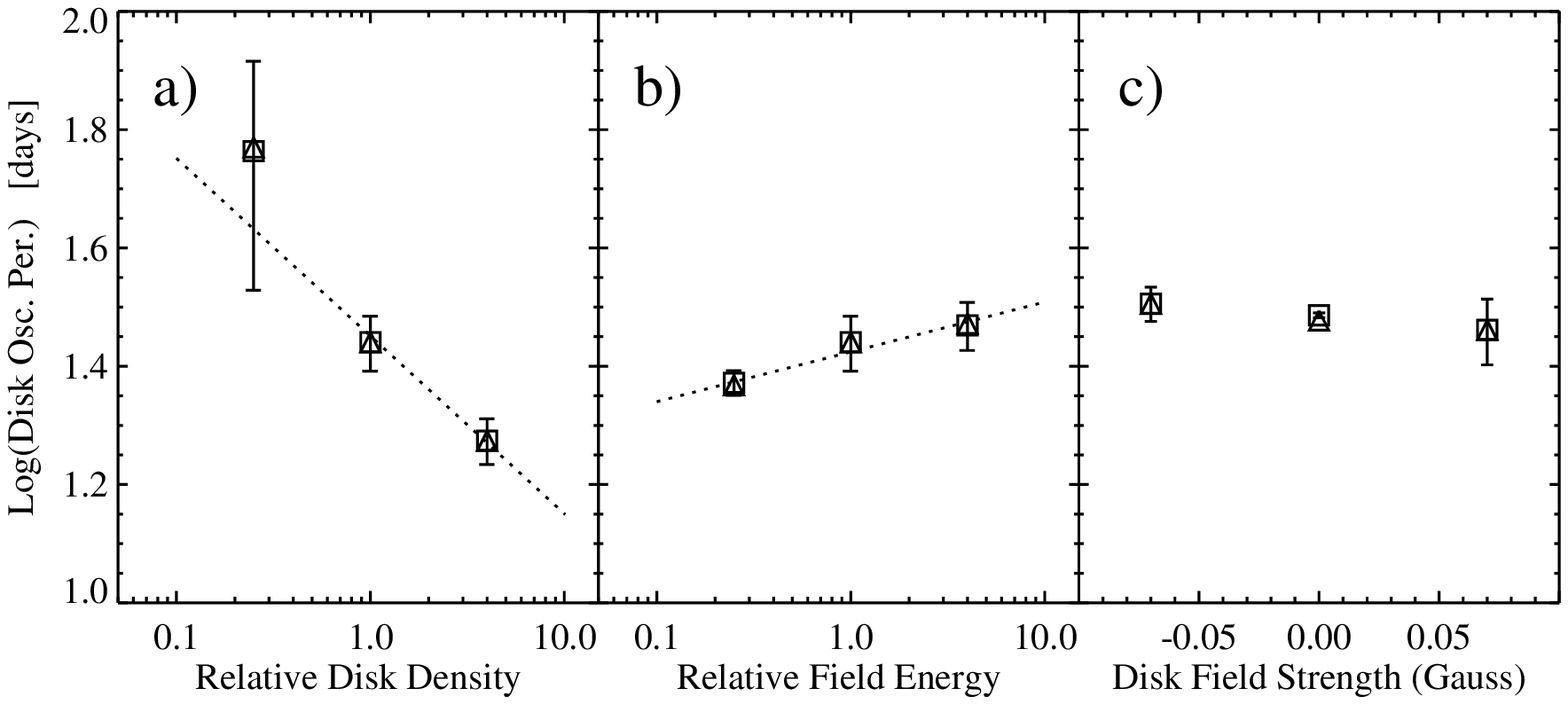}

\caption{The squares and dotted lines are as in figure \ref{maxvelfig}
but for disk oscillation period at $t$ = 160 days.  The triangles
represent a semi-analytical calculation (see text).
\label{periodfig}}

\end{figure*}

As seen in figures \ref{maxvelfig} -- \ref{luminosityfig},
\ref{jcompfig} -- \ref{periodfig}, and in table \ref{sensitivetab},
variation of key system parameters do indeed cause variations in
outflow properties, but the relationship is fairly weak (the strongest
power law relating a system property to the initial conditions has $n
= -0.44$).  This self-regulating nature of the system suggests that
the self-consistent interaction between the disk, outflow, and stellar
magnetosphere ``softens'' the effect of any single parameter change.
For example, we found in \S \ref{angmom} that the large scale magnetic
field structure rapidly evolves to a configuration that removes
angular momentum from the disk at a rate that depends most strongly on
the evolved field and much more weakly on other parameters.  Also,
when the stellar magnetic field is changed, the outer region of the
evolved field (which is important for the extraction of angular
momentum) has characteristics which only partially reflect the surface
value change.

The robust and self-regulating nature of this system imply that, if
such a mechanism does occur in astrophysical systems, it should indeed
be ubiquitous, and may actually be the prime mechanism for the
formation of non-relativistic astrophysical jets.  Unfortunately the
self-regulating nature also means that observable (or deduceable)
properties of outflows, such as the mass outflow rate or the rate of
angular momentum expelled from the system, will contain a limited
amount of information about the properties of the star-disk systems.

Figures \ref{maxvelfig} -- \ref{luminosityfig} and \ref{jcompfig} --
\ref{periodfig} also demonstrated with some certainty that the system
properties examined do not depend on the amount or direction of
magnetic flux that initially threads the disk.  For each system
property we examined, there is at least a weak dependence on the
initial stellar magnetic field energy, evident in panel b) of each of
these figures.  However, as seen in the c) panels, for the cases with
additional vertical field (and therefore additional magnetic energy in
the disk), there were no convincing trends.  Implicitly, therefore, it
is the interaction between the stellar magnetosphere and the inner
edge of the accretion disk that drives the outflow, as opposed to the
existence (via artificial initial conditions) of a dipole field that
threads the disk everywhere at $t$ = 0.  Note, however, that for
stronger vertical fields, magnetocentrifugal launching of disk plasma
may increase at larger radii (leading to an increased collimation of
the material launched from the disk inner edge).  Magnetized disk
models \citep[e.g.,][]{pudritzouyed97,camenzind97,kuwabaraea00} will
become important, independent from and in addition to the mechanism
presented here.  A system with a dynamically important stellar dipole
field and disk-associated field might produce an outflow with an
episodic/unsteady core (closest to the axis of rotation) and a
``cocoon'' of material originating from the disk.

We also evaluated and modified the disk oscillations model of GW (\S
\ref{evaluation}).  Using a semi-analytical method, we tested and
confirmed many of their basic assumptions, including the idea that the
oscillation period is set by spin down time of disk inner edge, not by
the inflation of magnetosphere.  Therefore, we demonstrated that the
basic physical mechanism for disk oscillations is understood.

There is still work to be done with the accretion/ejection mechanism
presented here in order to decide whether such a process occurs in
real astrophysical systems.  Future work is planned to determine the
importance of other properties of the system, though the qualitative
mechanism is valid for a wide range of parameters.  In particular, the
stellar rotation rate certainly influences the details of the outflow
characteristics, but the inflation of the magnetosphere will always
occur, even in the extreme case of a non-rotating star, due to the
differential rotation in the accretion disk.  Also, a more realistic
treatment of the accretion disk, possibly including the effects of the
disk viscosity or the magnetorotational instability
\citep{balbushawley91}, would affect the star-disk interaction through
the accretion rate ($\dot M_{\rm outer}$), and will thus change the
period and amplitude of the disk oscillations.  Future work will also
examine the system under true three-dimensional conditions.  The
effect of radiative cooling/heating may not be negligible, especially
considering that the coupling of the field to the gas is dependent on
the specific ionization state and composition of the circumstellar
matter.

The disk oscillation process, and therefore the jet launching
mechanism, is fundamentally driven by a diffusive process which is
poorly understood.  Recall that the mechanism requires magnetic
reconnection (a diffusive process with physics that are partially
understood; see \citealp{priestforbes00}) for accretion and
re-expansion.  GBW point out that this model is not particular
sensitive to the physics of magnetic reconnection, since regions of
oppositely directed magnetic field are driven together by the disk
oscillations.  However, this is only a partial solution to the
dilemma, since the disk oscillation mechanism itself is initiated by
the diffusion of stellar magnetospheric flux into the disk inner edge.

The timescale for knot formation predicted by this mechanism is a
factor of $\sim 15$ different than observed for HH 30, and the
rudimentary scaling laws offered by this parametric study do not offer
an obvious solution to the discrepancy.  The timescale predicted by
this work (several tens of days) cannot presently be resolved by
direct imaging (there would be several tens of knots in a typical
optical resolution element), so one cannot yet rule it out.  However,
even if there exists a short timescale present in astrophysical
systems that can be explained by this mechanism, there is then no
detailed theoretical prediction of the observed longer timescale.  A
primary focus of future work with this model will be the timescale for
disk oscillations/knot formation.

Another criticism of this work (as well as many other models; see \S
\ref{intro}) addresses the assumption that the central star has a
strong dipole magnetic field (dominating over other multipolar
components which may exist).  Detailed observations of the magnetic
fields in T Tauri stars \citep[see, e.g.,][]{safier98,johnskrullea99}
suggest that strong, ordered fields may not dominate; rather, the
stellar field topology appears to be time-variable and
non-axisymmetric.  Whether such a field can participate in the
mechanism described here (e.g., via non-axisymmetric magnetic flux
tubes connecting the star and disk) is not known.

\acknowledgments

We thank Frank Shu, Chris Johns-Krull, and Bo Reipurth for useful
discussion about the model and observed outflow properties.  We also
thank the anonymous referee for helpful suggestions that improved the
paper.  This research was supported by NSF grant AST 97-29096.



\clearpage

\begin{deluxetable}{llll}
\tablewidth{0pt}
\tablecaption{Case matrix for parametric study. \label{summarytab}}

\tablehead{
\colhead{Case Name} &
\colhead{Variation} &
\colhead{Physical Parm.} &
\colhead{Comments}
}
\startdata
Baseline          & None & N/A & See GBW \\
Low Density       & $\rho_{\rm disk} \div 4.0$ 
   & $\rho_{\rm mid}^a$ = 0.55 & \nodata \\
High Density      & $\rho_{\rm disk} \times 4.0$ 
   & $\rho_{\rm mid}^a$ = 8.83 & \nodata \\
Weak Field        & $B_* \div 2.0$
   & $B_*$(pole) = 900 G & \nodata \\
Strong Field      & $B_* \times 2.0$
   & $B_*$(pole) = 3600 G & \nodata \\
Truncated Disk    & Disk at 19 $R_*$ 
   & N/A & Applied to weak field case \\
Closed Magn.      & Vertical field closes 
   & Primordial field of & Used for Truncated disk \\
   & magnetosphere & 0.07 G in $-z$ direction & case only \\
Open Magn. & Vertical field opens
   & Primordial field of & As above but oppositely \\
   & magnetosphere & 0.07 G in +$z$ direction & directed primordial field \\

\enddata

$^a$ $\rho_{\rm mid}$ is the mass density in the disk midplane at 10
$R_\odot$ in units of 10$^{-10}$ g cm$^{-3}$.

\end{deluxetable}

\begin{deluxetable}{lllrlrl}
\tablewidth{0pt}
\tablecaption{Fit parameters of large scale magnetic field strength. 
\label{bpowertab}}
\tablehead{
\colhead{Case} &
\colhead{$B_{01}^a$} &
\colhead{$n_1$} &
\colhead{$B_{02}$(tot)$^a$} &
\colhead{$n_2$(tot)} &
\colhead{$B_{02}$(pol)$^a$} &
\colhead{$n_2$(pol)}
}
\startdata
Baseline    & 2010 $\pm$ 10 & 2.98 $\pm$ 0.02 & 750 $\pm$ 10 &
1.76 $\pm$ 0.01 & 840 $\pm$ 20 & 1.94 $\pm$ 0.01 \\

Low Dens.   & 1890 $\pm$ 20 & 2.82 $\pm$ 0.04 & 900 $\pm$ 30 &
1.88 $\pm$ 0.02 & 960 $\pm$ 30 & 1.98 $\pm$ 0.02 \\

High Dens.  & 1930 $\pm$ 10 & 2.92 $\pm$ 0.01 & 760 $\pm$ 10 &
1.75 $\pm$ 0.01 & 940 $\pm$ 30 & 2.01 $\pm$ 0.02 \\

Weak Field  & 1050 $\pm$ 10 & 2.89 $\pm$ 0.01 & 550 $\pm$ 10 &
1.71 $\pm$ 0.01 & 580 $\pm$ 20 & 1.88 $\pm$ 0.03 \\

Strong Field& 3800 $\pm$ 10 & 2.94 $\pm$ 0.01 & 1180$\pm$ 10 &
1.83 $\pm$ 0.01 & 1450$\pm$ 40 & 2.01 $\pm$ 0.02 \\

Trunc. Disk & 1090 $\pm$ 10 & 2.89 $\pm$ 0.01 & 550 $\pm$ 20 &
1.74 $\pm$ 0.01 & 560 $\pm$ 10 & 1.85 $\pm$ 0.02 \\

Closed Magn.& 1100 $\pm$ 10 & 2.90 $\pm$ 0.02 & 550 $\pm$ 10 &
1.73 $\pm$ 0.01 & 570 $\pm$ 10 & 1.88 $\pm$ 0.02 \\

Open Magn.  & 1100 $\pm$ 10 & 2.87 $\pm$ 0.02 & 550 $\pm$ 10 &
1.71 $\pm$ 0.01 & 560 $\pm$ 10 & 1.82 $\pm$ 0.02 \\

\enddata

$^a$ in Gauss

\end{deluxetable}

\begin{deluxetable}{lll}

\tablewidth{0pt}
\tablecaption{Summary of dependences.
\label{sensitivetab}}

\tablehead{
\colhead{Outflow} &
\colhead{Disk} &
\colhead{Initial Dipole} \\
\colhead{Property} &
\colhead{Density} &
\colhead{Magnetic Energy}
}

\startdata

Wind Morphology & Independent & Independent \\

Outflow Velocity   & $-0.24$ & $~~0.32$ \\

$\dot M_{\rm w}$   & $~~0.23 \pm 0.06$ & $~~0.22 \pm 0.10$ \\

$L_{\rm w}$        & $-0.05 \pm 0.06$ & $~~0.43 \pm 0.04$ \\

$\dot J_{\rm w}$   & $~~0.06 \pm 0.07$ & $~~0.30 \pm 0.12$ \\




Period$^a$	   & $-0.30 \pm 0.05$ & $~~0.08 \pm 0.02$ \\

Amplitude$^a$	   & $-0.44 \pm 0.09$ & $~~0.08 \pm 0.08$ \\

\enddata

$^a$ Refers to the oscillations of the disk inner edge, which drive
oscillations in the outflow.

\end{deluxetable}

\end{document}